\documentstyle[prl,aps,multicol,epsfig]{revtex}

\begin{document}

\def \be #1{\begin{equation}\label{#1}}
\def \ee {\end{equation}}
\def \bea #1{\begin{eqnarray}\label{#1}}
\def \eea {\end{eqnarray}}
\def \Eq #1{Eq.~(\ref{#1})}
\def \Fig #1{Fig.~\ref{#1}}
\def \F2 {FPL${}^2$ }
\def \D2 {DPL${}^2$ }
\def \Rs {\sf I\hskip-1.5pt R} 
\def \Zs {\mbox{\sf Z\hskip-5pt Z}} 
\def \Cs {\rm C\!\!\!I\:}
\def \bp {\mbox{\boldmath $\partial$}}
\def \rb {\rm b}
\def \rg {\rm g}
\def \Kt {{\sf I\hskip-1.5pt K}}

\title{Transition from the Compact to the Dense Phase of
Two-Dimensional Polymers}

\author{Jesper Lykke Jacobsen$^1$%
        \footnote{E-mail: lykke@lps.ens.fr; lykke@dfi.aau.dk}
        and Jan{\'e} Kondev$^{2,3}$%
        \footnote{E-mail: janek@ias.edu}}

\address{$^1$Laboratoire de Physique Statistique%
             \footnote{Laboratoire associ{\'e} aux universit{\'e}s
                       Paris 6, Paris 7 et au CNRS.},
             Ecole Normale Sup{\'e}rieure,
             24 rue Lhomond, 75231 Paris Cedex 05, France \\
         $^2$Institute for Advanced Study, Olden Lane, Princeton, NJ 08540 \\
         $^3$Department of Physics, Princeton University, Princeton, NJ 08540}
         
\date{\today}

\maketitle

\begin{abstract}
 We present a unifying picture of the compact, dense and dilute phases of
 two-dimensional polymers. The lattice dependence of the scaling exponents for 
 compact polymers is reconciled with their universality in the dense and 
 dilute case.
 In particular, we show that violations of the fully-packing constraint in
 the compact phase can be interpreted as magnetic screening in the
 associated Coulomb gas, which induces a flow to either the dense
 or the dilute phase. 
 When more than one flavour of polymers is present the flow away from the
 compact phase leads to a decoupling of the flavours, and the central
 charge decreases by an integer. If charge asymmetry develops the
 polymer flavours may independently flow to either of the two non-compact
 phases.
\end{abstract}


\begin{multicols}{2}

\section{Introduction}

Unlike  polymers in the much studied  dense and dilute phases \cite{Nienhuis},
two-dimensional polymers in the compact phase exhibit a remarkable
lack of universality. 
Namely, their scaling properties are  described by
conformal field theories that depend crucially on the coordination number
of the underlying lattice. This leads to an unusual situation where critical 
exponents for compact polymers on the square and the honeycomb lattice are 
different \cite{nien_fpl,batch94,jk_jpa,batch96,jj_npb}.

Another  intriguing property of  compact polymers, which by
definition cover {\em all} the lattice sites, is that they 
constitute new universality classes that differ from that of dense 
polymers \cite{DupSaleur87}, for which the fraction of covered sites $0<f<1$. 
In the dilute phase, which corresponds to a polymer in a good solvent, $f=0$.

The difference between compact and dense polymers is easily  understood from 
the random surface perspective. Namely, two-dimensional polymers can be 
mapped to a lattice model of a fluctuating 
interface which is described in the scaling limit by a Liouville field theory
with imaginary couplings. However, whilst the dense and dilute 
phases are both adequately described by a scalar height%
\footnote{When more than one loop flavour can be defined we shall, not
  surprisingly, find that one height component is needed for each of
  these flavours. However, unlike what is the case for compact
  polymers these height components decouple in the effective field
  theory.},
the fully-packing 
(or Hamiltonian) constraint imposed on compact polymers forces the height 
space to be  $(z-1)$-dimensional, $z$ being the coordination number of 
the lattice at hand.
The larger dimensionality of the height space leads to a more complicated
effective field theory and to new exponents \cite{jk_jpa,jj_npb}.  

Nonetheless, certain critical exponents turn out to be ``super-universal'',
in the sense that they are identical for dense polymers and 
compact polymers on the square and honeycomb lattice. 
Consider the exponents $x_s$, for 
$s=1,2,\ldots$, governing the power-law decay of the correlation function
$G_s(r) \sim r^{-2x_s}$ which describes 
$s$ polymer chains with one end anchored within a small region 
centred around the  origin, and the other in a region around a distant 
point ${\bf r}$.  A comparison of the explicit expressions for
these so-called {\em string} (or {\em watermelon}) dimensions for 
the three models under  consideration reveals that they are the  same
for $s$ even. This is also true of the difference 
$x_s - x_{s'}$ between any two odd string dimensions. These observations have
important consequences for the physically relevant
{\em contact exponents}%
\footnote{Globular proteins in their native state form compact structures,
  and hydrophobic interactions that take place at contacts play an 
  important role in the folding process \cite{chan_dill}.}
$\theta_{\cal G}$. These exponents determine the asymptotic decay of the 
probability $P_{\cal G}(y) \sim y^{\theta_{\cal G}}$ as $y \to 0$, 
that a certain number of points on the polymer separated by macroscopic 
distances {\em along} the chain, 
simultaneously have a spatial separation of order $y$; the arrangement of 
contact points is specified by the graph ${\cal G}$. 
Restricting for simplicity the attention to two-point
contacts, it is easily seen that $\theta_{\cal G}$ only depends on the string
dimensions through the universal combinations $x_3 - x_1$ and $x_4$, as
long as only interior points and at most one of the endpoints of the polymer 
are involved in the contact. On the other hand, contacts involving {\em both} 
endpoints give rise to exponents that depend on $x_1$ separately, and
hence are not 
universal. Incidentally, $x_1$ also determines the conformational exponent
$\gamma = 1-x_1$, which describes the scaling of the ratio of the number of 
open to closed polymer chains with their length, and is a measure of the 
effective entropic interaction between the two polymer ends. 
For compact polymers on the honeycomb lattice
the mean-field \cite{orland} result $x_1^{\rm ch} = 0$ was found
\cite{batch94,jk_jpa}, whereas both compact polymers on
the square lattice and dense polymers exhibit an effective entropic repulsion
between chain ends with $x_1^{\rm cs} = -5/112$ and $x_1^{\rm d} = -3/16$
respectively \cite{jj_npb,DupSaleur87}.

This paper aims at providing a theory of the transition from
the compact to the dense phase, with emphasis on the
mechanism that restores the universality of the string dimensions when 
violations of the fully-packing constraint are permitted. A particularly
interesting framework for these investigations is an extension of the
square-lattice model which was used to solve the compact polymer problem 
in Ref.~\cite{jj_npb}.
Namely, this model allows for the definition of {\em two} mutually excluding
flavours (species) of polymers, which in the compact case were found to
interact in a manner that endowed the model with a two-dimensional manifold
of critical fixed points. Apart from the obvious interest in examining what
happens to the Liouville field theory as one flows away from the
compact phase, it is a most beguiling question whether the dense and dilute
phases also allow for a two-flavour extension. We shall find the answer to
this question to be affirmative, with an important caveat: Once vertices are 
allowed to be uncovered by the polymers, the two flavours {\em decouple}. 
Quite remarkably, this decoupling turns out to be so complete that one of 
the flavours may even flow to the
dilute phase whilst the other remains dense! This provides a rationalisation
for four of the five critical branches of the O($n$) model studied in 
Ref.~\cite{Blote89}.

We begin (in Section~\ref{sec:plasma}) by giving a heuristic argument,
based on the field theory for the 
compact phase, that explains how dense exponents may emerge from the compact
ones when we allow for violations of the $f=1$ constraint. In the Coulomb gas
formalism string dimensions such as the ones defined above are determined 
by inserting a pair of electromagnetic test charges into the vacuum, 
and calculating  their interaction energy. 
Violations of the compactness constraint are shown to correspond to 
magnetic charges which form a plasma that completely screens {\em one}
of the components of the vector magnetic 
charge of the test particles. As a result the compact string dimensions
reduce to those of two non-interacting flavours of dense polymers.

Though physically appealing this argument is  unsound. 
We therefore go on (in Section \ref{sec:DPL2}) to
propose a lattice model that 
exhibits all three phases of polymers---compact, dense, and dilute.
Exact values of the central charge and the
critical exponents are then computed from a Coulomb gas construction 
which, although not mathematically rigorous, is well founded in the 
renormalisation group \cite{Nienhuis}. These
results neatly confirm the more intuitive magnetic plasma picture, and go 
further in describing the interplay between the dense and dilute phase.

Even though the two-flavoured model on the square lattice is our prime
concern, the ideas presented here are quite general. In particular
they apply to the one-flavoured model on the honeycomb lattice, and
in general to loop models defined on arbitrary $z$-fold
coordinated regular lattices. This is the subject of the Discussion
(Section \ref{sec:discussion}),
where we also elaborate on the relation between our results and those
obtained for the O($n$) model of Ref.~\cite{Blote89}.

\section{Coulomb gas with a magnetic plasma}
\label{sec:plasma}

\subsection{Two-flavour fully packed loop model}
\label{sec:FPL2}

In Ref.~\cite{jj_npb} scaling properties of compact polymers on the square
lattice were investigated with the help of the two-flavour fully
packed loop (FPL${}^2$) model. This model has a two-dimensional region of
critical fixed points wherein  compact polymers are identified with a 
single point; this point also belongs to a line 
of fixed points which is a subset of the critical manifold and it 
describes interacting compact polymers.   
Since our results for the compact-to-dense transition extend to the entire
critical manifold we shall begin by briefly reviewing the FPL${}^2$ model.

First define ${\cal G'}$ to be the set of all {\em edge colourings}
of the square lattice with four different colours. In other words,
each edge of the  lattice is assigned a colour 
${\bf A}$, ${\bf B}$, ${\bf C}$ or ${\bf D}$, subject to the constraint that
for any vertex all of its four adjacant edges carry different colours. 
If we bipartition the square lattice into even and odd sublattices, 
directed {\em loops} of two
different flavours (``black'' and ``grey'') can be defined as follows:
Colour ${\bf A}$ designates a black loop segment directed from an even to
an odd site, and colour ${\bf B}$ corresponds to a black loop segment having
the opposite direction. Similarly colours ${\bf C}$ and ${\bf D}$ are used to
define directed grey loops. By ${\cal G}$ we denote the equivalence classes
of ${\cal G'}$ with respect to independent changes of directions for each of
the loops. The FPL$^2$ partition function is then
\begin{equation}
  Z = \sum_{\cal G} n_{\rm b}^{N_{\rm b}} n_{\rm g}^{N_{\rm g}} \ ,
  \label{FPL2}
\end{equation}
where $n_{\rm b},n_{\rm g}$ are the fugacities and $N_{\rm b},N_{\rm g}$ the
numbers of (undirected) black and grey loops in ${\cal G}$ respectively. The
model (\ref{FPL2}) is critical on the manifold
$0 \le n_{\rm b}, n_{\rm g} \le 2$,
and the compact polymer problem discussed in the Introduction is recovered
in the limit $(n_{\rm b},n_{\rm g}) \to (0,1)$.

The extra orientational information present
in ${\cal G'}$ allows for a local redistribution of the loop fugacities.
Indeed, assign a phase factor $\exp(\pm {\rm i}\pi e_{\rm b}/4)$ to each
vertex where a directed black loop makes a right (left) turn, and the factor
$1$ if it continues straight. In this way the weight of the entire black
loop becomes $\exp(\pm {\rm i}\pi e_{\rm b})$ if the orientation is
clockwise (counter-clockwise). The undirected loop weight $n_{\rm b}$ is
obtained by summing over the two orientations, whence
$n_{\rm b} = 2 \cos(\pi e_{\rm b})$. With a similar convention for the grey
loops, and letting $\lambda({\bf x})$ be the product of the local black
and grey weights at the vertex ${\bf x}$, the partition function, Eq.~(\ref{FPL2}), 
can now be rewritten as
\begin{equation}
  Z = \sum_{{\cal G'}} \prod_{{\bf x}} \lambda ({\bf x}) \ .
  \label{vertex-weights}
\end{equation}

The continuum limit of the FPL$^2$ model is obtained by mapping the oriented
loop configurations to an interface model defined on the dual lattice. The
microscopic heights ${\bf z}$ live on the plaquettes of the square lattice 
and they are defined, up to an overall shift, by the edge colours which 
stipulate the height differences between neighbouring plaquettes.
More precisely, when encircling an even (odd) site in the clockwise direction
a vector ${\bf A}$, ${\bf B}$, ${\bf C}$ or ${\bf D}$ is
added to (subtracted from) ${\bf z}$ depending on the state of the edge being
crossed. The fully-packing constraint imposes the condition
\begin{equation}
  {\bf A} + {\bf B} + {\bf C} + {\bf D} = {\bf 0} \ ,
  \label{constraint1}
\end{equation}
whence the colour vectors are three-dimensional.
A suitable representation is
\begin{eqnarray}
  {\bf A} = (-1,+1,+1) \ , \ \ \ \ {\bf B} = (+1,+1,-1) \ , \nonumber \\
  {\bf C} = (-1,-1,-1) \ , \ \ \ \ {\bf D} = (+1,-1,+1) \ .
  \label{colours1}
\end{eqnarray}
The continuum limit is then taken by coarse graining ${\bf z}$. 
The effective field theory is described by a Liouville action
containing three terms \cite{jk_prl}:
\begin{equation}
  S = S_{\rm E} + S_{\rm B} + S_{\rm L} \ .
  \label{action}
\end{equation}
The elastic part $S_{\rm E}$ is a gradient-squared term in the coarse grained
height ${\bf h}({\bf x})$, and it controls the fluctuations of the interface
away from flat configurations. The boundary term $S_{\rm B}$ assigns the 
correct
weights to loops winding around the point at infinity by coupling ${\bf h}$
to the scalar curvature $\tilde{\cal R}$. Finally, $S_{\rm L}$ is a Liouville
potential that assigns the correct weights to loops in the bulk, and it is 
the coarse grained version of the microscopic vertex weights 
$\lambda({\bf x})$.

In the dual 
Coulomb gas picture the action, Eq.~(\ref{action}), is interpreted as that of
a two-dimensional gas of electro-magnetic vector charges interacting via a
logarithmic potential. Electric charges are associated with vertex operators
$\exp({\rm i} {\bf e} \cdot {\bf h})$ which  appear as the scaling limits
of FPL$^2$ operators that can be expressed as local, periodic functions of
the microscopic heights. Magnetic charges ${\bf m}$ act as non-local 
constraints on the height field, and they are associated with topological 
defects in the microscopic heights or, equivalently, with violations
of the four-colouring constraint. Finally, 
the curvature term $S_{\rm B}$ introduces a background electric charge
$-2{\bf e}_0$ at the boundary of the system.  For reasons of neutrality
this endows the Coulomb gas vacuum with a floating charge $+2{\bf e}_0$ located
in the bulk of the system.

\subsection{Height defects and magnetic screening}

Following the standard Coulomb gas recipe \cite{Nienhuis}, the
calculation of the scaling dimension of an electromagnetic operator
reduces to the determination of the interaction energy 
of ``test particles'' of charge $\pm ({\bf e},{\bf m})$ 
inserted into the vacuum. This energy varies logarithmically with separation,
where the prefactor of the logarithm is twice the sought dimension. 
As the floating charge is free to move around the bulk it will coalesce with 
one of the test particles thus minimising the energy.  
Therefore  the interaction
energy consists of three terms associated with the pairwise interactions of 
the background electric charge $(-2{\bf e}_0, 0)$,
$({\bf e},{\bf m})$, and
$(2{\bf e}_0 - {\bf e},-{\bf m})$, with the vacuum energy subtracted off.  
The energy of the vacuum is simply due to the pair interaction between the
background charge and the floating charge. 

Consider, for example, the computation of the string dimensions $x_s$
for the FPL$^2$ model. Since the defect strings may now have any of
the two flavour labels this generalises to $x_{s_{\rm b},s_{\rm g}}$,
where $s_{\rm b}$ ($s_{\rm g}$) denotes the number of black (grey)
strings. For dislocations generated in the bulk the number
$s_{\rm b} + s_{\rm g}$ must be even \cite{jj_npb}.
Appropriate defect configurations may be specified through the colouring
state of the four edges around some fixed vertex \cite{jk_prb}. For instance,
$({\bf A},{\bf A},{\bf C},{\bf D})$ generates two directed black strings
and corresponds to a height vortex of strength
${\bf m}_{2,0} = {\bf A} - {\bf B} = (-2,0,2)$. Similarly,
$({\bf A},{\bf B},{\bf C},{\bf C})$ is a vortex of strength
${\bf m}_{0,2} = {\bf C} - {\bf D} = (-2,0,-2)$ that generates two grey
strings, and $({\bf A},{\bf C},{\bf C},{\bf D})$ generates one string of
either flavour and has strength
${\bf m}_{1,1} = {\bf C} - {\bf B} = (-2,-2,0)$.
For simplicity we shall concentrate on the black strings and define
$x_s = x_{s,s_{\rm g}}$, where $s_{\rm g} = s \mbox{ mod } 2$. The defect
magnetic charge
\begin{equation}
  {\bf m}_s = \left \lbrace
  \begin{array}{ll}
    (-s,0,s)      & \mbox{for $s$ even,} \\
    (-s-1,-2,s-1) & \mbox{for $s$ odd,}
  \end{array}
  \right.
  \label{m_s}
\end{equation}
found as a suitable linear combination of ${\bf m}_{2,0}$ and
${\bf m}_{1,1}$, must be accompanied by an appropriate compensating electric
charge that takes care of the spurious phase factors arising from the winding
of the strings around the defect cores \cite{Nienhuis}.
Setting the fugacity of the grey strings equal to unity ($e_{\rm g} = 1/3$)
we obtain the following result for the scaling dimensions:
\begin{equation}
  2 x_s = \left \lbrace
  \begin{array}{ll}
    \frac14 (1-e_{\rm b}) s^2 - \frac{e_{\rm b}^2}{1-e_{\rm b}} &
    \mbox{for $s$ even,} \\
    \frac14 (1-e_{\rm b}) s^2 - \frac{e_{\rm b}^2}{1-e_{\rm b}} +
    \frac{2(1-e_{\rm b})}{5-3e_{\rm b}} &
    \mbox{for $s$ odd.}
  \end{array}
  \right.
  \label{x_compact}
\end{equation}
As mentioned in the Introduction the even scaling dimensions agree with
those of dense polymers \cite{DupSaleur87} (and also with those of the
fully-packed loop model on the honeycomb lattice \cite{jk_jpa}), whereas the
odd dimensions contain an extra term due to the coupling between the black
and the grey strings.

So far our discussion has been for the fully-packed case where a fraction
$f=1$ of the lattice vertices are visited by both loop flavours. Now consider
moving infinitesimally into the dense phase by diminishing the fraction of
visited vertices to $f=1-\epsilon$. One can imagine doing so by distributing
a fraction $\epsilon$ of $({\bf A},{\bf B},{\bf A},{\bf B})$ defects and
a fraction $\epsilon$ of $({\bf C},{\bf D},{\bf C},{\bf D})$ defects
throughout the system. The corresponding vortices have magnetic charge
$\pm {\bf m}_T$, where
\begin{equation}
  {\bf m}_T = (0,4,0) \ .
  \label{m_T}
\end{equation}
Physically this situation corresponds to having a dilute magnetic plasma
superimposed on the Coulomb gas vacuum.

As before we can then imagine inserting two electromagnetic test particles
of charge $\pm ({\bf e},{\bf m})$ into the system in order to measure the
string scaling dimensions. Just as in the familiar case of electric test
particles in a dielectric medium there will now be some screening, the
difference being that here it is the magnetic charge that is being screened.
However, unlike what is the case in this dielectric analogy, 
the magnetic plasma is completely free to move around, and the screening
will be complete. This is a consequence of the fact that the magnetic charges
in \Eq{m_T} are {\em relevant}, i.e. their scaling dimension \cite{jj_npb}
\be{dim_T}
 2 x(\pm {\bf m}_T) =
 4 \ \frac{(1-e_{\rb})(1-e_{\rg})}{(1-e_{\rb})+(1-e_{\rg})} 
\ee
is less than 
2  on the whole critical manifold of the \F2 model. Therefore these charges
are not bound into vortex-antivortex pairs at large scales \cite{KT}. 

Taking our cue from this intuitive picture we hypothesise that the
effect of the magnetic plasma can be modeled by the substitution
\begin{equation}
  {\bf m}  \equiv (m^1,m^2,m^3) \longrightarrow
  {\bf m}' \equiv (m^1,0,m^3) \ ,
\end{equation}
since the only non-vanishing component of the screening charge (\ref{m_T})
is along the 2-direction, and this does not couple to any of the other
directions. In particular, replacing ${\bf m}_s$ of Eq.~(\ref{m_s}) by
${\bf m}_s'$ we find that instead of Eq.~(\ref{x_compact}) we get
\begin{equation}
  2 x_s' = \frac14 (1-e_{\rm b}) s^2 - \frac{e_{\rm b}^2}{1-e_{\rm b}}
\end{equation}
for all $s$, whether even or odd! It seems that the interaction between
the two loop flavours that was responsible for the difference between the
even and odd string dimensions in Eq.~(\ref{x_compact}) has somehow been
disposed of, and what remains is the critical exponents of {\em dense} 
polymers \cite{DupSaleur87}.

One can proceed to compute the central charge for the screened system.
We recall that in the compact ($f=1$) case this was given as
$c = 3 + 12x({\bf e}_0,{\bf 0})$, where the three free bosons present
in $S_{\rm E}$ each contribute unit central charge, and there is a negative
shift due to the background electric charge. However, since the magnetic
charges live in height space the complete screening of $m^2$ means that
height fluctuations along the 2-direction are frozen out. This is nothing 
but the usual Kosterlitz-Thouless scenario \cite{KT},
where due to the unbinding of dislocations the 
interface defined by the second height component is rendered smooth. As a 
result the central charge for $f = 1 - \epsilon$ is
$c'= c - 1 = 2 + 12x({\bf e}_0,{\bf 0}')$, which can be written as
\begin{equation}
  c'(e_{\rm b},e_{\rm g}) = c^{\rm dn}(e_{\rm b}) + c^{\rm dn}(e_{\rm g}) \ ,
  \label{sumrule}
\end{equation}
where
\begin{equation}
  c^{\rm dn}(e) = 1 - \frac{6 e^2}{1-e}
\end{equation}
is nothing but the central charge of one flavour of dense polymers
\cite{DupSaleur87}. If we again set the fugacity of the grey loops equal to
unity ($e_{\rm g} = 1/3$) we find that $c^{\rm dn}(e_{\rm g}) = 0$, and only
the contribution from the black loops remains.
The sum rule (\ref{sumrule}) should be taken as a further
indication that once we allow for violations of the fully-packing constraint
the two loop flavours decouple.

At this point, of course, a number of objections may be raised.
First, it is seen from Eq.~(\ref{m_s}) that the magnitude of the magnetic
charge along the 2-direction is $-2$. However, for the screening to take
place it then seems that the magnetic screening charges $\pm {\bf m}_T$
given by Eq.~(\ref{m_T}) must somehow be fractionalised.

Another objection is that the above calculation, when taken at face value,
purports to be perturbative around the fully-packed phase. However, as 
pointed out earlier [\Eq{dim_T}] the defect operators 
within the FPL$^2$ model are strongly relevant! This means that the 
fugacity of these defects will flow off to infinity at large
scales. Hence, the magnetic plasma analogy is no more than a
heuristic argument. Nevertheless, we shall see in the following section
that it is possible to generalise the FPL$^2$ model to a lattice model
that explicitly accommodates configurations that are not fully packed.
In this generalised model Coulomb gas calculations can be carried
out, confirming the main results found in this Section.

Even when accepting the magnetic plasma analogy as a heuristic framework
for treating the case of $f<1$, the FPL$^2$ height mapping on which it is
based breaks down completely for $f \to 0$. It is therefore impossible to
say anything about the role of the dilute polymer phase. However, from the
calculations presented in the next Section we shall be able to say more about
how dilute exponents may emerge.

\section{Two-flavour densely packed loop model}
\label{sec:DPL2}

In order to unify our understanding of compact, dense and dilute polymers
as well as their possible interrelation we introduce a  
statistical mechanics model that can accommodate all three  phases. 
Since the critical properties of the compact phase are known to 
be lattice dependent we need to consider
the model on a specific lattice. To examine the important issues associated
with the presence of more than one loop flavour we choose the square
lattice, whereas the question of what the construction would look like
on other lattices is deferred to the Discussion.

\begin{figure}
  \noindent\begin{minipage}{8.66cm}
   \epsfxsize=8.6cm \epsfbox{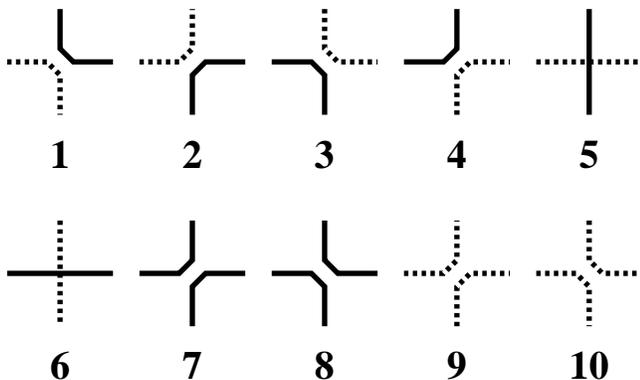}
   \protect\caption{\label{fig:vertices}The ten allowed vertices in the
    DPL$^2$ model. Black and grey loop segments are represented by solid
    and dashed linestyle respectively.}
  \end{minipage}
\end{figure}

We therefore define the {\em two-flavour densely packed loop} (DPL$^2$)
{\em model} on the square lattice by listing its ten allowed vertex states in
Fig.~\ref{fig:vertices}. The first six vertices are those familiar from
the fully packed (FPL$^2$) case, whilst the last four vertices explicitly
allow the model to violate the fully-packing constraint. Vertices 7--8
exclude the grey loops from a given site and carry a weight $W_{\rm b}$,
and similarly vertices 9--10 exclude the black loops and have weight
$W_{\rm g}$. The first six vertices are all assigned unit weight.
The partition function of the DPL$^2$ model is then
\begin{equation}
  Z = \sum_{{\cal G}^{\rm d}} W_{\rm b}^{v_{\rm b}} W_{\rm g}^{v_{\rm g}}
                              n_{\rm b}^{N_{\rm b}} n_{\rm g}^{N_{\rm g}} \ ,
  \label{DPL2}
\end{equation}
where $n_{\rm b},n_{\rm g},N_{\rm b},N_{\rm g}$ have the same meaning as
in Eq.~(\ref{FPL2}), and $v_{\rm b}$ and $v_{\rm g}$ are the number of
occurrences of vertices 7--8 and 9--10, respectively. The summation runs
over the set ${\cal G}^{\rm d}$ of loop configurations on the square
lattice that can be made out of vertices 1--10, subject to the constraint
that both flavours form closed loops. This model is closely related to the
O($n$) model introduced in Ref.~\cite{Blote89} for which five branches of
critical points were found as a function of $n$. In the Discussion we shall
comment on the precise relation of the \D2 model to this one, thus providing
a physical picture of its critical properties.   

Clearly, in the special case $(W_{\rm b},W_{\rm g}) = (0,0)$ the DPL$^2$ model
reduces to the FPL$^2$ model, and the analysis of Ref.~\cite{jj_npb} applies.
Let us briefly recall the main results of this fully packed case. 
For $s_{\rm b}+s_{\rm g}=0 \mbox{ mod } 2$ the full spectrum of string
dimensions is
\begin{eqnarray}
 2 x_{s_{\rb}, s_{\rg}} &=&
     \frac14 \left[ (1-e_{\rm b}) s_{\rb}^2 +
     (1-e_{\rm g}) s_{\rg}^2 \right] \nonumber \\
 &-& \left[ \frac{e_{\rm b}^2}{1-e_{\rm b}}(1-\delta_{s_{\rb},0}) +
     \frac{e_{\rm g}^2}{1-e_{\rm g}}(1-\delta_{s_{\rg},0}) \right]
 \label{x-FPL2}  \\
 &+& \delta^{(2)}_{s_{\rb},1} \delta^{(2)}_{s_{\rg},1}
     \frac{(1-e_{\rm b})(1-e_{\rm g})}{(1-e_{\rm b})+(1-e_{\rm g})} \nonumber,
\end{eqnarray}
where $\delta^{(2)}_{i,j} \equiv \delta_{i=j \rm{(mod \ 2)}}$.
The result for the central charge is
\begin{equation}
  c = 3 - 6\left( \frac{e_{\rm b}^2}{1 - e_{\rm b}}
      + \frac{e_{\rm g}^2}{1 - e_{\rm g}} \right) \ .
  \label{c-FPL2}
\end{equation}

In the following we focus on the Coulomb gas construction for the case where
either $W_{\rm b}$ or $W_{\rm g}$ is greater than zero.

\subsection{Colouring and height representation}

The mapping from oriented loop configurations ${\cal G'}^{\rm d}$ to
the four-colouring representation goes through exactly as described for the
FPL$^2$ model in Sec.~\ref{sec:FPL2}. When defining the microscopic
heights ${\bf z}$ there is however one very important
difference. The presence of vertices 7--10 in Fig.~\ref{fig:vertices}
imposes extra constraints on the choice of the colour vectors:
\begin{equation}
  {\bf A} + {\bf B} = {\bf 0} \ , \ \ \ \
  {\bf C} + {\bf D} = {\bf 0} \ .
  \label{constraint2}
\end{equation}
Actually, whenever one of these constraints holds true the other follows
from Eq.~(\ref{constraint1}). So the conditions (\ref{constraint2}) are
valid provided that either $W_{\rm b} > 0$ or $W_{\rm g} > 0$.

As a consequence the color vectors in Eq.~(\ref{colours1}) must
be replaced by
\begin{eqnarray}
  {\bf A} = (+1,0), \ \ \ \ {\bf B} = (-1,0), \nonumber \\
  {\bf C} = (0,+1), \ \ \ \ {\bf D} = (0,-1),
  \label{colours2}
\end{eqnarray}
{\em i.e.}, the height space is now two-dimensional. This elimination of one
of the height components is very reminiscent of the effect of screening  
in the magnetic plasma analogy of Sec.~\ref{sec:plasma}.

The local redistribution of the loop fugacities, $n_{\rm b}$ and $n_{\rm g}$ of
Eq.~(\ref{DPL2}), in terms of local weight factors associated with the left
and right turns of the loops, again takes place as in Sec.~\ref{sec:FPL2}.
One may wonder how should the vertex weights
$W_{\rm b}$ and $W_{\rm g}$ be represented? 
Unlike the loop weights which do not renormalise under a change of scale%
\footnote{This is a consequence of the {\em loop ansatz} \cite{jk_prl,jj_npb}
  to which we shall return in Section \ref{sec:loop-ansatz} below.},
the vertex weights
do change under scale transformations as they flow towards their fixed-point
(and non-universal) values. Thus, from the point of view of the Liouville
field theory description that we are about to set up for the scaling limit, 
they need not be represented at all%
\footnote{Actually, this is not quite true. As mentioned above the constraints
 (\ref{constraint2}) explicitly take into account that at least one of
 $W_{\rm b}$ or $W_{\rm g}$ is non-zero. Furthermore, one may suspect that
 if the bare value of $W_{\rm b}$ is very large the grey loops flow to the
 dilute phase. This is indeed the case, and we shall see below how this
 mechanism is implemented in the Coulomb gas construction.}!
Of course, all along our working assumption is that the scaling limit of the 
\D2 model exists  for loop fugacities $0\le n_{\rb}, n_{\rg}\le 2$ .

\subsection{Ideal states}

An ideal state of a loop model has two distinguishing features.
First, it is macroscopically flat, so that fluctuations of the interface
around the ideal states can be controlled by an elastic term $S_{\rm E}$
in the effective action. Second, it is entropically selected in the sense 
that it maximises the number of configurations which differ from it 
by the smallest allowed  change. To be concrete consider
the directed FPL$^2$ model in the four-colouring representation. 
Since any change of a configuration  must necessarily involve
the change of some edge colour $\sigma_1$ into another colour $\sigma_2$,
it follows that the whole loop
$\sigma_1 \sigma_2 \sigma_1 \sigma_2 \cdots \sigma_1 \sigma_2$
defined as an alternating sequence of these two colours must be changed into
$\sigma_2 \sigma_1 \sigma_2 \sigma_1 \cdots \sigma_2 \sigma_1$.
An ideal state allows for the maximum number of such {\em loop flips},
and consequently has a maximum number of small loops of alternating colour.

\begin{figure}
  \noindent\begin{minipage}{8.66cm}
   \epsfxsize=8.6cm \epsfbox{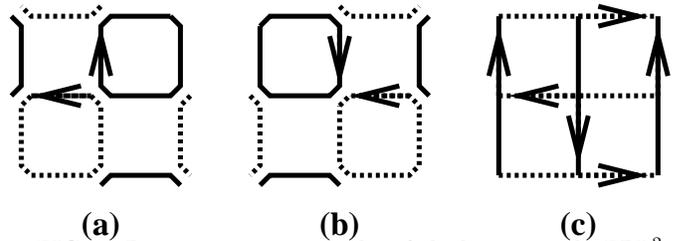}
   \protect\caption{\label{fig:id-states}Representative examples of ideal
    states in the FPL$^2$ model, here shown in the oriented loop
    representation. The full ideal state is obtained by tiling $\Rs^2$
    with the $2 \times 2$ pattern given in the figure.
    Each of the examples corresponds to a group of eight ideal states,
    related by independent changes of orientations for the two flavours,
    and by flavour conjugation.}
  \end{minipage}
\end{figure}

In the FPL$^2$ model on the square lattice there are 24 ideal states,
defined by the 4! possible permutations of the colours ${\bf A}$, ${\bf B}$,
${\bf C}$ and ${\bf D}$ around a fixed vertex. For later convenience we
choose to label them by the clockwise sequence 
$(\sigma_1,\sigma_2,\sigma_3,\sigma_4)$ of colours around this vertex,
starting with the leftmost edge. In the oriented loop representation the
ideal states form a pattern of $2 \times 2$
plaquettes which is repeated throughout the lattice. Three representative
examples are shown in Fig.~\ref{fig:id-states}.a--c. Taking into account the
possible changes of loop flavours and directions there are eight FPL$^2$
ideal states corresponding to each of these examples, making a total of 24. 

\begin{figure}
  \noindent\begin{minipage}{8.66cm}
   \epsfxsize=8.6cm \epsfbox{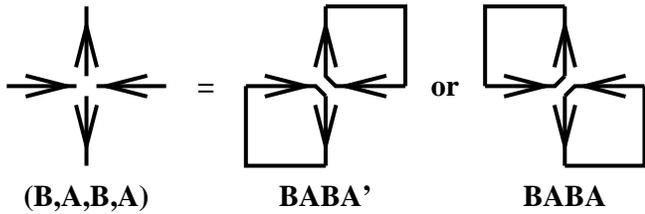}
   \protect\caption{\label{fig:ambiguity}Certain colour configurations around
    a vertex do not uniquely specify the directed loop configuration. In
    particular the label $({\bf B},{\bf A},{\bf B},{\bf A})$ corresponds to
    two distinct DPL$^2$ ideal states, which we distinguish by a prime as
    shown.}
  \end{minipage}
\end{figure}

When we allow for the extra vertices of the DPL$^2$ model more ideal states
become possible. Namely, for each of the 16 states exemplified by
Fig.~\ref{fig:id-states}.a--b there corresponds another ideal state
where only one of the two loop flavours is used to form the small loops.
Again we can label the states by means of the colour configuration around
a fixed vertex, but as shown in Fig.~\ref{fig:ambiguity} this is in some
cases no longer unique. To remedy this we adopt the convention to add a prime
to the label whenever the loop configuration around a vertex is as shown in
Fig.~\ref{fig:id-states}.a; see Fig.~\ref{fig:ambiguity}. 

This ambiguity has also bearings on the way in which changes of a given
colouring configuration may be performed. Namely, when changing some edge
colour $\sigma_1$ into $\sigma_2$ one may encounter one of the ambiguous
vertices of Fig.~\ref{fig:ambiguity} in the process of tracing out the loop
$\sigma_1 \sigma_2 \sigma_1 \sigma_2 \cdots \sigma_1 \sigma_2$
to be flipped. In that case one has two
equivalent choices for proceding. However, this peculiarity does not
change the fact that the 16 new states are still ideal, since they are
perfectly macroscopically flat and they allow for an equal number of small
changes as do the 24 ideal states of the FPL$^2$ model. The DPL$^2$ model
thus has 40 ideal states.

The lesson to be learned from this complication is that it is favourable to
think of the DPL$^2$ ideal states in terms of the oriented loop representation
rather than in terms of the colouring representation. 
In fact, the mapping
from oriented loop configurations to the four-colouring representation is
now many-to-one, so that the inverse mapping is no longer well-defined. Since
we only need the colours to define the heights,  which are the basic
constituents of the continuum field theory,  we do not have to worry whether
the series of mappings can be reversed.

\subsection{More on the dimensionality of height space}

It is worthwhile noticing that the choice (\ref{colours2}) would have been
just as good as (\ref{colours1}) in ensuring the one-to-one correspondence
between oriented \underline{F}PL$^2$ configurations and the
microscopic heights. 
However, a too restrictive choice of the colour vectors may render a
non-ideal state macroscopically flat, in which case the elastic term
$S_{\rm E}$ of the action fails to enforce the proper entropic penalty
for a state that does not allow for a maximum number of loop flips.

\begin{figure}
  \noindent\begin{minipage}{8.66cm}
   \epsfxsize=8.6cm \epsfbox{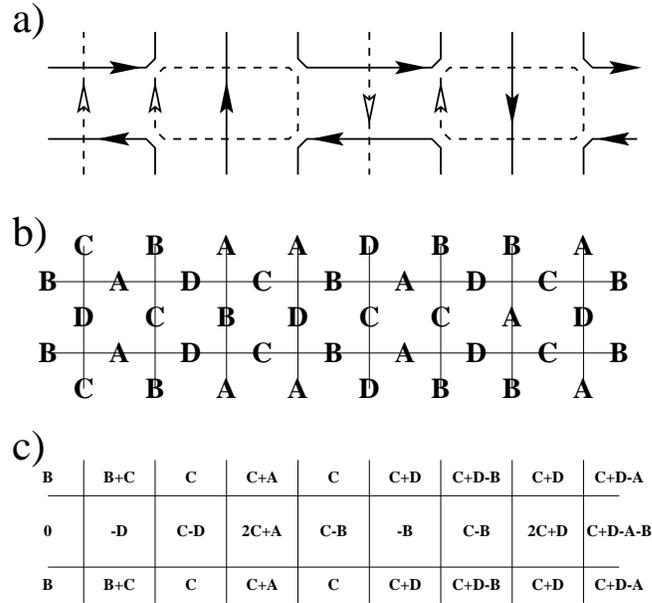}
   \protect\caption{\label{fig:spooky-state}This FPL$^2$ state has a
   non-vanishing macroscopic height gradient 
   along the horizontal direction. It is shown in three equivalent
   representations: (a) oriented loops, (b) edge colourings, and (c)
   microscopic heights.}
  \end{minipage}
\end{figure}

As an example consider the state arising from a tiling of $\Rs^2$ with
the $2 \times 8$ pattern shown in Fig.~\ref{fig:spooky-state}. Upon
coarse graining this state is flat in the vertical ($x^2$) direction,
but has a macroscopic height gradient in the horizontal direction:
\begin{equation}
  \partial {\bf h} / \partial x^1 =
  \frac18 \: [({\bf C} + {\bf D}) - ({\bf A} + {\bf B})] \ .
\end{equation}
With the choice (\ref{colours2}) the height gradient vanishes, and
$S_{\rm E} = 0$ for this height configuration.
On the other hand, the choice (\ref{colours1}) leads to
$\partial {\bf h} / \partial x^1 = (0,-1/2,0)$, and the state is duly
suppressed by $S_{\rm E}$ since the stiffness constant along the
second height direction%
\footnote{The similarity between this expression and Eq.~(\ref{dim_T})
  for the thermal scaling dimension emphasises the special role of the
  second height component.}
\begin{equation}
  K_{22} = \frac{(1-e_{\rm b})(1-e_{\rm g})}
                {(1-e_{\rm b}) + (1-e_{\rm g})}
\end{equation}
is non-zero throughout the critical region \cite{jj_npb}.

This important point is further illuminated by noting that black and
grey loops play the role of contour lines for the projection of ${\bf h}$ 
along the $(1,0,-1)$ and the $(1,0,1)$ directions
respectively. Evidently these projections fail to detect any variation
along the 2-direction in height space, and accordingly the state given
by Fig.~\ref{fig:spooky-state} {\em appears} macroscopically flat when
viewed exclusively in terms of its oriented loop representation.

\subsection{Ideal state graph}
\label{sec:id-graph}

The 40 ideal states of the DPL$^2$ model define the {\em ideal state graph}
${\cal I}$ which is instrumental in the construction of the continuum field
theory. The idea is to identify the ideal states with nodes on ${\cal I}$
whose position is given by the average microscopic height
$\langle {\bf z} \rangle$ of the given state. Two ideal states are said to
be nearest neighbours if their respective microscopic heights are identical
on three fourths of the lattice plaquettes. On ${\cal I}$, the two nodes
corresponding to such a nearest neighbour pair are connected through a link.

\begin{figure}
  \noindent\begin{minipage}{8.66cm}
   \epsfxsize=8.6cm \epsfbox{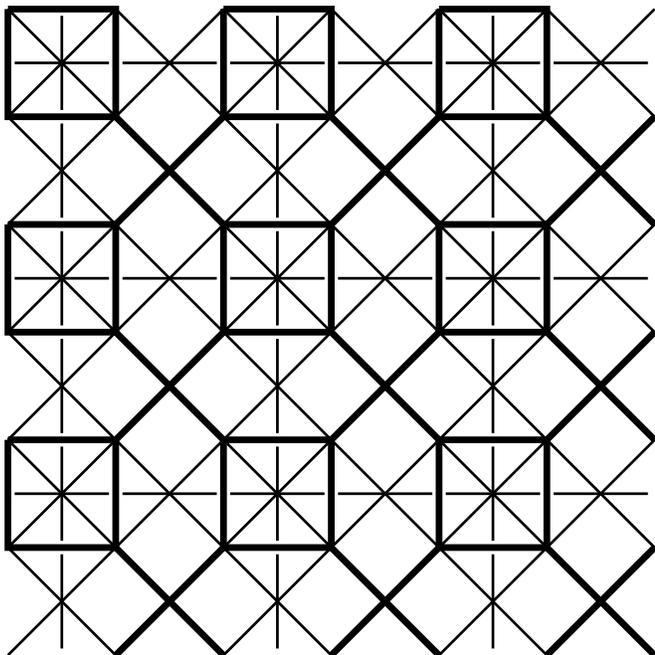}
   \protect\caption{\label{fig:ideal}Ideal state graph ${\cal I}$ of the
    DPL$^2$ model. The thick and thin linestyles distinguish the links
    which are inherited from the FPL$^2$ model from those which are new
    to the DPL$^2$ model. The scale is set by the squares in thick
    linestyle, which have side length $1/2$.}
  \end{minipage}
\end{figure}

The result of this construction is the ideal state graph shown in
Fig.~\ref{fig:ideal}. Here the links added due to the inclusion of the
16 ideal states 
which are new to the DPL$^2$ model are shown as thin lines, whereas the
thick linestyle designates links between the 24 ideal states which have
been inherited from the FPL$^2$ model. We recall that the ideal state
graph for the FPL$^2$ model was a covering of $\Rs^3$ by truncated octahedra.
Since the DPL$^2$ colour vectors (\ref{colours2}) can be obtained from their
FPL$^2$ counterparts (\ref{colours1}) by a projection onto the
$z^1z^3$-plane and a subsequent $\frac{3\pi}{4}$ rotation and rescaling,
it should hardly come as a surprise that the thick lines form  a
two-dimensional projection of the truncated octahedra. In particular, the
square and the hexagonal faces are clearly visible.

\begin{figure}[h]
  \noindent\begin{minipage}{8.66cm}
   \epsfxsize=8.6cm \epsfbox{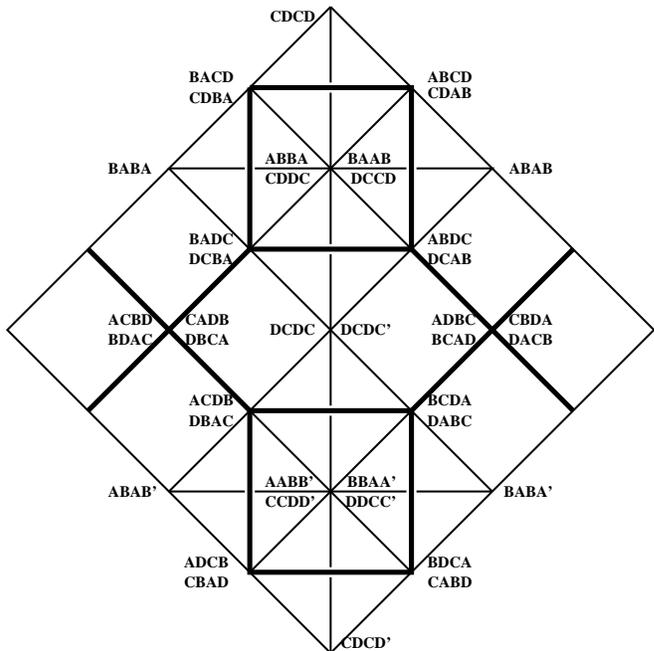}
   \protect\caption{\label{fig:repeat}The repeating unit of ${\cal I}$ along
     with the state labels. Ideal states that are new to the DPL$^2$
     model can easily be distinguished by noting that only two different
     colour vectors appear in their state label.}
  \end{minipage}
\end{figure}

As a consequence of our definition of nearest neighbour ideal states any
one ideal state corresponds to an infinity of nodes in ${\cal I}$. The
entire ideal state graph thus corresponds to a covering of $\Rs^2$ by the
repeating unit shown in Fig.~\ref{fig:repeat}. Two novel features of the
DPL$^2$ ideal state graph which have not been encountered in the previously
studied loop models are that each node may have several labels, and that
links may cross without being connected by a node. The latter observation
just indicates that ${\cal I}$ really {\em must} be thought of as a graph
(a collection of nodes and links) and not simply a lattice; of course, unlike
the usual notion of a graph, the precise location of the nodes is of paramount
importance as it bears on the effective field theory of the model.

Concentrating first on the 24 ideal states on Fig.~\ref{fig:repeat} that are
connected by thick links the division of the states into the three classes
exemplified by Fig.~\ref{fig:id-states} becomes obvious. The two groups of
eight states having the elementary plaquette loops positioned as on
Fig.~\ref{fig:id-states}.a and Fig.~\ref{fig:id-states}.b respectively, 
each form a square with doubly labeled nodes. These squares are connected
by thick lines via
two four-fold labeled nodes (``crosses'' on Fig.~\ref{fig:repeat}) that
together accommodate the eight states of the type shown in
Fig.~\ref{fig:id-states}.c.%
\footnote{At first sight these states hardly appear ``ideal'', at least
  not when viewed solely in the oriented loop representation. However,
  it should be clear that they are needed to ensure the connectivity
  of ${\cal I}$, {\em cfr.}~Fig.~\ref{fig:ideal}.}
The remaining 16 ideal states which are new
to the DPL$^2$ model, located at intersections of thin lines 
in Fig.~\ref{fig:repeat}, occur in
two groups of eight states corresponding to plaquette loops positioned 
as in Fig.~\ref{fig:id-states}.a (primed states) and in
Fig.~\ref{fig:id-states}.b (unprimed states). Quite naturally, the primed
states are linked to the square of eight FPL$^2$ states with loops
positioned as shown in Fig.~\ref{fig:id-states}.a, and similarly the unprimed
states are distributed around the other square
({\em cfr.}~Fig.~\ref{fig:id-states}.b). In particular, the states
${\bf DCDC}$ and ${\bf DCDC}'$ must be thought of as labeling two
{\em distinct} nodes, which just happen to occupy identical positions
on Fig.~\ref{fig:repeat}.%
\footnote{A similar remark holds true, of course, for the primed and the
 unprimed versions of states ${\bf ABAB}$, ${\bf BABA}$ and ${\bf CDCD}$.}
This subtlety becomes more evident if one tries to trace out ${\cal I}$
explicitly by moving from one ideal state to the neighbouring one. This 
is accomplished by either choosing two colours in the starting ideal
state and then interchanging them along all the alternatingly coloured
loops to obtain a new state, or by changing one loop flavour to the other
(this leads to a new ideal state only for states 
that are made up of all gray or all black loops).  It then turns out to be 
impossible to move between states on
two different squares using only the links that are shown as thin lines
on Fig.~\ref{fig:repeat}.

\subsection{Liouville field theory}

A continuum field theory for the DPL$^2$ model can now be obtained by
coarse graining the microscopic heights over domains of ideal states
\cite{jk_npb}. When defining the continuum height field, nodes in
${\cal I}$ representing the same ideal state should be identified,
which means that the height must be compactified with respect to the
{\em repeat lattice} ${\cal R}$,
\begin{equation}
  {\bf h}({\bf x}) \in \Rs^2 / {\cal R} \ .
\end{equation}
According to Fig.~\ref{fig:repeat} the repeat lattice is a square lattice
of side $\sqrt{2}$, spanned by the vectors $(1,1)$ and $(1,-1)$.

We are now ready to write down the Liouville field theory for the
height field ${\bf h}({\bf x})$. As usual the partition function $Z_>$, which 
describes  only the large-scale fluctuations of the height, can be written 
as a functional integral
\begin{eqnarray}
  Z_> &=& \int {\cal D}{\bf h} \, \exp(-S[{\bf h}]), \nonumber \\
  S   &=& S_{\rm E} + S_{\rm B} + S_{\rm L},
  \label{LFT}
\end{eqnarray}
where the effective Euclidean action consists of three terms that were 
mentioned in section \ref{sec:FPL2}. Here we discuss them in some detail.

\subsubsection{Elastic term}
\label{sec:elastic}

The most general form of the elastic term has the form
\begin{equation}
  S_{\rm E} = \frac12 \int {\rm d}^2{\bf x} \,
              K_{\alpha \beta}^{ij} \partial_i h^{\alpha} \partial_j h^{\beta},
\end{equation}
where Latin letters label the basal plane coordinates and Greek letters
pertain to the height space. Since, by definition of the allowed DPL$^2$
vertices (see Fig.~\ref{fig:vertices}), the continuum model is invariant
with respect to rotations in the basal plane, the stiffness tensor is
diagonal in the Latin indices:
$K_{\alpha \beta}^{ij} = K_{\alpha \beta} \delta^{ij}$.
The remaining four components are constrained by colour symmetries,
corresponding to independent reversals of the two flavours of directed loops:
\begin{equation}
  \begin{array}{ll}
  {\bf A} \leftrightarrow {\bf B}: &
  e_{\rm b} \leftrightarrow -e_{\rm b} \mbox{ and } z^1 \leftrightarrow -z^1 \\
  {\bf C} \leftrightarrow {\bf D}: &
  e_{\rm g} \leftrightarrow -e_{\rm g} \mbox{ and } z^2 \leftrightarrow -z^2 \ .
  \end{array}
\end{equation}
Any one of these symmetries prevents the term
$K_{12} \bp h^1 \cdot \bp h^2$ from occurring in $S_{\rm E}$.
The elastic term thus assumes the diagonal form
\begin{equation}
  S_{\rm E} = \frac12 \int {\rm d}^2{\bf x} \, \left[
              K_1 (\bp h^1)^2 + K_2 (\bp h^2)^2 \right] \ ,
\end{equation}
involving only two independent stiffness constants $K_1$ and $K_2$.

\subsubsection{Boundary term}

The boundary term $S_{\rm B}$ generically takes on the form
\begin{equation}
  S_{\rm B} = \frac{{\rm i}}{4 \pi} \int {\rm d}^2{\bf x}
              ({\bf e}_0 \cdot {\bf h}) \tilde{\cal R} \ ,
\end{equation}
where $\tilde{\cal R}$ is the scalar curvature, and ${\bf e}_0$ is the
background electric charge which is to be determined.
To calculate ${\bf e}_0$ it is most convenient to 
consider the DPL$^2$ model defined on a cylinder,
so that the curvature is concentrated at either end:
\begin{equation}
  \tilde{\cal R} = 4\pi [\delta(+\infty)-\delta(-\infty)] \ .
\end{equation}
This terms in the action has the effect of placing two vertex operators
with charges $\pm {\bf e}_0$ at the far ends of the cylinder. They in 
turn give complex weights to loops winding around
the cylinder. In order for these to be the same as those defined earlier
for loops in the bulk, the equations
\begin{eqnarray}
  {\bf e}_0 \cdot {\bf A} = \pi e_{\rm b} \ , \ \ \ \
  {\bf e}_0 \cdot {\bf B} = -\pi e_{\rm b} \ , \nonumber \\
  {\bf e}_0 \cdot {\bf C} = \pi e_{\rm g} \ , \ \ \ \
  {\bf e}_0 \cdot {\bf D} = -\pi e_{\rm g} \ ,
\end{eqnarray}
must be satisfied. The unique solution is 
\begin{equation}
  {\bf e}_0 = (\pi e_{\rm b},\pi e_{\rm g}) \ .
\end{equation}

\subsubsection{Liouville potential}

According to Eqs.~(\ref{vertex-weights}) and (\ref{LFT})
the appropriate form of the Liouville potential is
\begin{equation}
  S_{\rm L} = \int {\rm d}^2{\bf x} \, w[{\bf h}({\bf x})] \ ,
\end{equation}
where $\exp(-w[{\bf h}({\bf x})])$ is the scaling limit of the vertex
weights $\lambda({\bf x})$ arising from the local redistribution of the
loop fugacities $n_{\rm b}$ and $n_{\rm g}$. Since these are uniform in
each of the ideal states we can consider $w({\bf h})$ to be a function of
the coarse grained height ${\bf h} \in {\cal I}$.

It is reassuring to check that whenever a node in ${\cal I}$ carries
multiple labels ({\em cfr}.~Fig.~\ref{fig:repeat}) all of the concerned
states have the same value of the vertex weight $\lambda({\bf x})$.
The reason is that the corresponding oriented loop configurations
either (a) are connected by a translation in the basal plane, or (b)
have a local cancellation of left and right turns, or (c) all carry
unit weight. The only apparent exception is pairs of states of the type
${\bf DCDC}$, ${\bf DCDC}'$. These pairs have already been discussed in
Sec.~\ref{sec:id-graph}, where we found that the two states should
really be associated with two {\em distinct} nodes, which just happen
to be spatially coincident. The fact that
$w({\bf DCDC}) = + {\rm i}\pi e_{\rm g}/2$, whereas
$w({\bf DCDC}') = - {\rm i}\pi e_{\rm g}/2$
therefore does not lead to any contradiction.

Any local function of the colours (operator) which is uniform in the
ideal states defines a periodic function of ${\bf h} \in {\cal I}$
with the periods forming the repeat lattice ${\cal R}$. Upon coarse
graining of the height this periodicity  property is conserved. 
Therefore the continuum limit of any such operator can be expanded as a
Fourier series, 
where the basis functions in the expansion are the vertex operators
$\exp({\rm i}{\bf e} \cdot {\bf h}({\bf x}))$. Consequently, 
\begin{equation}
  w[{\bf h}({\bf x})] = 
    \sum_{{\bf e}\in{\cal R}^*_w} \tilde{w}_{\bf e} \exp({\rm i}
    {\bf e} \cdot {\bf h}({\bf x})) \ .
  \label{Fourier}
\end{equation}
Note that the sum does not run over the reciprocal lattice ${\cal R}^*$
but rather over some sublattice ${\cal R}_w^* \subset {\cal R}^*$. This is
because $w({\bf h})$ may have some additional periodicity within the
unit cell of ${\cal I}$  shown in Fig.~\ref{fig:repeat}. A careful analysis
reveals that this is indeed the case: The height periods of $w({\bf h})$
form a square lattice spanned by $(1,0)$ and $(0,1)$, and so ${\cal R}_w^*$
is a square lattice of side $2\pi$, spanned by $(2\pi,0)$ and
$(0,2\pi)$.

\subsubsection{Loop ansatz}
\label{sec:loop-ansatz}

In order to relate the stiffness constants $K_1$ and $K_2$ to $e_{\rm b}$
and $e_{\rm g}$ we need only determine the most relevant vertex operators in
the expansion (\ref{Fourier}) and impose the condition that they be exactly
marginal. This is the {\em loop ansatz} which ensures that loop weights
do not renormalise under a change of scale \cite{jk_prl}. From the viewpoint of
conformal field theory the electric charges appearing in these most relevant
vertex operators play the role of  the screening charges introduced by 
Dotsenko and Fateev \cite{Dotsenko}.

The scaling dimension of a general electromagnetic operator is given by
\cite{Dotsenko,jj_npb}
\begin{equation}
  2x({\bf e},{\bf m}) = \frac{1}{2\pi} \left[
    {\bf e} \cdot {\bf \sf K}^{-1} \cdot ({\bf e} - 2{\bf e}_0) +
    {\bf m} \cdot {\bf \sf K} \cdot {\bf m} \right] \ ,
  \label{em_dim}
\end{equation}
where the stiffness tensor ${\bf \sf K}$ according to Sec.~\ref{sec:elastic}
assumes the form
\begin{equation}
  {\bf \sf K} = \left[ \begin{array}{cc} K_1 & 0 \\ 0 & K_2 \end{array} \right] \ .
\end{equation}
The candidates for the screening charges are the four shortest vectors
in ${\cal R}_w^*$
\begin{eqnarray}
  {\bf e}_w^{(1)} = (+2\pi,0) \ , \ \ \ \
  {\bf e}_w^{(2)} = (-2\pi,0) \ , \nonumber \\
  {\bf e}_w^{(3)} = (0,+2\pi) \ , \ \ \ \
  {\bf e}_w^{(4)} = (0,-2\pi) \ ,
  \label{screenings}
\end{eqnarray}
and the scaling dimensions of the corresponding vertex operators are
\begin{eqnarray}
  x({\bf e}_w^{(1)},{\bf 0}) = \frac{\pi(1-e_{\rm b})}{K_1} \ , \ \ \ \
  x({\bf e}_w^{(2)},{\bf 0}) = \frac{\pi(1+e_{\rm b})}{K_1} \ , \nonumber \\
  x({\bf e}_w^{(3)},{\bf 0}) = \frac{\pi(1-e_{\rm b})}{K_2} \ , \ \ \ \
  x({\bf e}_w^{(4)},{\bf 0}) = \frac{\pi(1+e_{\rm b})}{K_2} \ .
\end{eqnarray}
At $(n_{\rm b},n_{\rm g}) = (2,2)$ the background electric charge vanishes
and the four operators are pairwise degenerate. However, for a generic
point on the critical manifold $0 \le n_{\rm b},n_{\rm g} \le 2$ the
operators $\exp({\rm i} {\bf e}_w^{(1)} \cdot {\bf h}({\bf x}))$ and
$\exp({\rm i} {\bf e}_w^{(3)} \cdot {\bf h}({\bf x}))$ are more relevant
than the other two.

The loop ansatz now leads to the result
\begin{equation}
  K_1 = \frac{\pi}{2} (1-e_{\rm b}) \ , \ \ \ \
  K_2 = \frac{\pi}{2} (1-e_{\rm g}) \ .
\end{equation}
In Sec.~\ref{sec:exponents} we shall see that this choice for the elastic
constants reproduces the critical exponents for two non-interacting
flavours of dense polymers.

\subsection{Charge asymmetry and the dilute phase}

In the case of a Coulomb gas with scalar electric charges it has been shown
by Nienhuis \cite{Nienhuis} that for particular values of the bare
particle fugacities a charge asymmetry may evolve, so that the renormalised
Coulomb gas contains unit charges of one sign only. Unit charges of the
opposite sign then appear only as excitations, having as a result that the
amplitude of the most relevant term in their two-point correlator vanishes.

An inspection of Nienhuis' argument reveals that this mechanism is rather
general, and in particular it may occur in the case of electromagnetic
vector charges that we are concerned with here. In our Liouville field
theory approach the vanishing of the above-mentioned amplitude translates
into the possibility of having one or more vanishing expansion coefficients
$\tilde{w}_{\bf e}$ in Eq.~(\ref{Fourier}).

Obviously it would be quite difficult to determine for which values of
the bare vertex weights $W_{\rm b}$ and $W_{\rm g}$ such a charge asymmetry
may evolve. Here we shall just mention that the four candidates for the
screening charges given in Eq.~(\ref{screenings}) in general warrant the 
following possibilities for the elastic constants
\begin{equation}
  K_1 = \frac{\pi}{2}(1 \mp e_{\rm b}) \ , \ \ \ \
  K_2 = \frac{\pi}{2}(1 \mp e_{\rm g}) \ .
\label{stiff_val}
\end{equation}
As already mentioned the upper signs will lead to the critical exponents
of dense polymers, and they correspond to the charge symmetric case. 
A remarkable observation is that the lower signs similarly
reproduce the critical exponents of two decoupled flavours of {\em dilute}
polymers%
\footnote{A similar phenomenon is observed in the Liouville field theory
 solution of the O$(n)$ model \cite{jk_unpub}.}.
Furthermore, nothing seems to prevent a charge asymmetry from
developing which would select the upper sign for
one of the couplings and the lower one for the other. Thus, the
complete decoupling of the loop flavours allows for a situation where
they each reside in either of the two non-compact phases ({\em i.e.},
dense or dilute).

\subsection{Symmetry algebra}

The \F2 model at the point $(n_{\rb},n_{\rg})$ = $(2,2)$ was shown 
to possess, in the continuum limit, an $su(4)$ affine Lie algebra symmetry,  
at level $k=1$ \cite{jk_npb}. This follows from the  effective field theory of 
the model which is  given by a three component height (three free 
massless bosons) compactified on the {\em root lattice} of $su(4)$.
(As in any conformal field theory there are actually two copies of 
the symmetry  algebra, associated with the holomorphic and the antiholomorphic 
components of the height field.) 
Yet another view of what transpires when the vertices that 
allow for the violation of the fully packing constraint are included, is
provided by the question: What happens to the symmetry algebra? 

In order to address this question we examine the
\D2 model at the point $(n_{\rb},n_{\rg})$ = $(2,2)$. In previous sections 
it was shown that the effective field theory of this model is given by a 
two-component height with the two components completely decoupled. From the 
analysis of the ideal state graph it was further shown that each component is  
compactified on the one dimensional lattice ${\cal R} = \sqrt{2} \Zs$. 
The combination of the lattice constant and the calculated stiffness
$K\equiv K_1=K_2 = \pi/2$ (\Eq{stiff_val}) is such that this lattice can
be identified with the root  lattice of an $su(2)$ algebra. 
Therefore, the \D2 model at the special point is characterised in the
scaling limit by an $su(2)_{k=1} \oplus su(2)_{k=1}$ affine Lie
algebra. The heights make up  the free field representation of the
appropriate Wess-Zumino-Witten model.

The generators of the holomorphic half of the 
symmetry algebra are the modes of the holomorphic  
currents \cite{fuchs}. There is a total of six currents, {\em i.e.}, 
conformal dimension $(1,0)$ operators, three for  
each of the two $su(2)$ algebras: 
\be{currents1}
 {\rm i} \partial h^1(z), \ \ \exp({\rm i} {\bf e}_w^{(1)}\cdot {\bf h}), \ \ 
 \ \exp({\rm i} {\bf e}_w^{(2)}\cdot {\bf h}(z))
\ee
and 
\be{currents2}
 {\rm i} \partial h^2(z), \ \ \exp({\rm i} {\bf e}_w^{(3)}\cdot {\bf h}(z)), \ \ 
 \ \exp({\rm i} {\bf e}_w^{(4)}\cdot {\bf h}(z)) \ ,  
\ee
where the electric charges appearing in the above vertex operators are 
the screening charges of  \Eq{screenings}. 
Here $z=x^1+{\rm i}x^2$ is the complex coordinate, whilst ${\bf h}(z)$ is
the holomorphic component of the height field, i.e.,
${\bf h}(x^1,x^2)= {\bf h}(z) + \bar{\bf h}(\bar{z})$.         

Finally we can give a concise answer to the question posed at the beginning of
this Section: The symmetry algebra of the compact phase is lowered from 
$su(4)$ to $su(2)\oplus su(2)$, as it flows to the dense phase. One of the 
three free bosons of $su(4)$ becomes massive, whilst the other two each form a 
free field representation of an $su(2)_{k=1}$ affine Lie algebra.

\subsection{Critical exponents}
\label{sec:exponents}

Having expressed the elastic constants in terms of the loop fugacities,
Eq.~(\ref{stiff_val}), we are now in the position to compute the central charge 
and the critical 
exponents of the DPL$^2$ model in the critical region $0\le n_{\rb}, 
n_{\rg} \le 2$.  

\subsubsection{Central charge}

For $(n_{\rm b},n_{\rm g}) = (2,2)$ the background charge vanishes, and
the Liouville theory (\ref{LFT}) consists only of the elastic term
$S_{\rm E}$. This is nothing but the action of two free bosons, yielding
a central charge of 2. In the general case there will be a shift due to
the background electric charge \cite{Dotsenko},
\begin{eqnarray}
  c &=& 2 + 12 x({\bf e}_0,{\bf 0}) \nonumber \\
    &=& 2 - 6\left( \frac{e_{\rm b}^2}{1 \mp e_{\rm b}}
          + \frac{e_{\rm g}^2}{1 \mp e_{\rm g}} \right).
\end{eqnarray}

If one chooses the upper signs this result agrees with that of the magnetic
plasma analogy, and according to Eq.~(\ref{sumrule}) it is consistent with
a situation where the two loop flavours have decoupled and are both  
in the dense phase of the O($n$) loop model. More generally we can write
\begin{equation}
  c(e_{\rm b},e_{\rm g}) =
  c^{\rm dn,dl}(e_{\rm b}) + c^{\rm dn,dl}(e_{\rm g}) \ ,
\label{c_D2}
\end{equation}
where $c^{\rm dn,dl}(e) = 1 - 6e^2/(1 \mp e)$ is the central charge of a single 
flavour of dense (dilute) loops.

\subsubsection{Geometrical scaling exponents}

As discussed in Sec.~\ref{sec:plasma} the complete spectrum of string
dimensions may be computed within the interface representation from the
two-point correlation functions between height defects. Using the
normalisation of Eq.~(\ref{colours2}) the Burgers charge of the 
elementary vortices is now given by
\begin{equation}
  {\bf m}_{2,0} = (2,0) \ , \ \ \ \
  {\bf m}_{0,2} = (0,2) \ , \ \ \ \
  {\bf m}_{1,1} = (1,1) \ .
\end{equation}
Furthermore, taking into account the compensating electric charge needed to
correct the spurious phase factors arising from the strings' winding
around the defect cores, the electromagnetic charge of a general string
defect generating $s_{\rm b}$ black and $s_{\rm g}$ grey strings becomes
\begin{equation}
  [{\bf e},{\bf m}]_{s_{\rm b},s_{\rm g}} =
  [(\pi e_{\rm b}(1-\delta_{s_{\rm b},0}),
    \pi e_{\rm g}(1-\delta_{s_{\rm g},0})) \, , \, (s_{\rm b},s_{\rm g})].
\end{equation}
We note that there is now {\em no} distinction between an even and an odd
number of strings.

The string dimensions, obtained from Eq.~(\ref{em_dim}), can finally be
written as
\begin{equation}
  x_{s_{\rm b},s_{\rm g}}(e_{\rm b},e_{\rm g}) =
    x_{s_{\rm b}}^{\rm dn,dl}(e_{\rm b}) +
    x_{s_{\rm g}}^{\rm dn,dl}(e_{\rm g}) \ ,
\label{string_D2}
\end{equation}
where
\begin{equation}
  2x_s^{\rm dn,dl}(e) = \frac14 (1 \mp e) s^2 -
                          \frac{e^2}{1 \mp e} (1 - \delta_{s,0})
\label{string_dndl}
\end{equation}
is the formula for  the string dimensions for one flavour 
of dense (dilute) O($n$)-loops.

\section{Discussion}
\label{sec:discussion}

In this paper we have presented a general mechanism by which compact polymers
may flow to the dense and dilute phases once violations of the fully-packing
constraint are allowed. The universal dense exponents were retrieved from the
lattice-dependent compact 
ones through a heuristic analogy to a magnetic plasma, or alternatively by
a Coulomb gas solution of a statistical mechanics model (the DPL$^2$
model) where two adjustable parameters ($W_{\rm b}$ and $W_{\rm g}$) were
used to control the deviations from fully packing. We have found that the
compact phase is unstable to any such deviation, however small, and that the
flow may be to the dense or the dilute phase depending on the bare values
of the adjustable parameters. Furthermore, by this mechanism the two loop
flavours of the FPL$^2$ model decouple completely, in the sense that they
may even flow to two distinct non-compact phases (dense or dilute). 
In terms of the Liouville field theory of the FPL$^2$ model, one of 
the three bosons becomes massive and the other two decouple.

\subsection{O($n$) model}

A particular variant of the O($n$) model on the square lattice 
was defined by Bl\"{o}te and Nienhuis \cite{Blote89}. Here the $n$-component 
spins live on the bonds of the square lattice whilst  two and four-spin 
interactions are defined for spins that share a common vertex. 
This spin model  has a graphical (loop) representation defined by 
the same vertices as the \D2 model. The only difference is in the 
assignment of vertex  weights, and, more importantly, the grey bonds are 
simply treated as being empty. 
This O($n$)  model is therefore a single-flavour loop model. We 
identify the O($n$) loops with the black loops of the \D2 model 
whose weight is $n_{\rb}=n$. The weight of    
the \D2 grey loops, which are formed by the empty bonds in the O($n$) model, 
is $n_{\rg}=1$ ({\em i.e.}, the weight of an O($n$) loop state  does
not depend on the number of loops formed by the empty bonds).  

If we focus for a moment on the central charge of the \D2 model along the line 
$n_{\rg}=1$, it is equal to the central charge of the black loops plus 
the central charge of the grey loops; see \Eq{c_D2}. The contribution of the 
grey loops to the central charge is $0$ or $1/2$ depending 
on whether they are in the dense or the dilute phase; this, of course, 
is determined by the choice of vertex weights. 
As the black loops can also be either in the dense or dilute phase this 
leads to four lines of critical points along which the central charge 
varies continuously. 

Going back to the O($n$) model of Bl\"{o}te and Nienhuis, these authors
identified, using numerical transfer matrix techniques,
{\em five} lines of critical fixed points with exponents continuously varying 
with $n$, each specified by a particular choice of the vertex weights.
One critical line (branch 0) was directly  
mapped to the critical $(n+1)^2$-state Potts model, since along this
line the flavour-crossing vertices 5 and 6 (see \Fig{fig:vertices})
are excluded. Based on the numerical
evidence (the measured  central charge and a few of the scaling
dimensions) two of the critical lines  were identified with the usual
dilute and dense phase of the O($n$) loop model (branch 1 and 2),  
one as a superposition of the dense O($n$) loop model and a
{\em critical Ising model} (branch 4), whilst for the 
final fifth line (branch 3) the numerics were inconclusive. 

{}From the above analysis of the central charge it is clear that
branches 1 and 2
can be identified with the grey loops being in the dense phase whilst the 
black loops are in the dilute and dense phase respectively. Along
branch 4 the black loops are dense whilst the gray loops are dilute. This is 
also confirmed by comparing our predictions for the dimension
$x_{1,1}$ ($x_H$ in \cite{Blote89}), Eqs.~(\ref{string_D2})
and (\ref{string_dndl}),  
with the numerically determined values in Ref.~\cite{Blote89}. 
We suspect that the ``anomalous'' branch 3 corresponds to both  
loop flavours being in the dilute phase.
For $n\ge 1$ this 
is born out by the numerics, but the correspondence seems to break down for
smaller values of the black loop weight. Why this is so 
remains  an interesting open question.

\subsection{Honeycomb lattice model}

We have mentioned earlier that the non-universality of the compact exponents
forced us to base our theoretical considerations on a particular lattice
model. In order to examine the effects of having more than one flavour of
loops we have opted for the square lattice. It is however reassuring to
see that an altogether similar construction can be built starting from the
honeycomb lattice, for which the fully-packed loop (FPL) model has 
been solved using Bethe ansatz methods \cite{batch94}.

The Coulomb gas construction for the honeycomb FPL model is based on
the three-colouring of the lattice edges \cite{jk_jpa}. The colour vectors
can be chosen as
\begin{equation}
  \begin{array}{ll}
  {\bf A} = \left( \frac{1}{\sqrt{3}} , 0       \right), & \\
  {\bf B} = \left(-\frac{1}{2\sqrt{3}}, \frac12 \right), &
  {\bf C} = \left(-\frac{1}{2\sqrt{3}},-\frac12 \right) \ ,
  \end{array}
  \label{3-colour}
\end{equation}
and oriented loops are defined in the usual way as alternating sequences
of colours ${\bf B}$ and ${\bf C}$.
In the corresponding densely-packed loop (DPL) model, where the vertex
$({\bf A},{\bf A},{\bf A})$ is allowed and violates the three-colouring
constraint, the height field becomes one-dimensional:
\begin{equation}
  {\bf A} = 0, \ \ \ \
  {\bf B} = 1, \ \ \ \
  {\bf C} = -1 \ .
\end{equation}
This is consistent with the magnetic plasma analogy, according to which
the proliferation of defect charges ${\bf m} = 3{\bf A} = (3,0)$ would
lead to the complete screening of the first height component in
Eq.~(\ref{3-colour}).

\begin{figure}
  \noindent\begin{minipage}{8.66cm}
   \epsfxsize=8.6cm \epsfbox{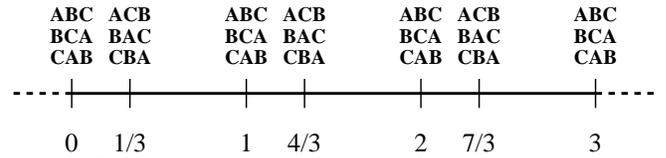}
   \protect\caption{\label{fig:DPL-ideal}Ideal state graph ${\cal I}$ of the
     DPL model on the honeycomb lattice. The repeat lattice is seen to be
     ${\cal R} = \Zs$.}
  \end{minipage}
\end{figure}

The FPL and the DPL models have the same six ideal states, and by
projection onto the 2-direction in height space 
the DPL ideal state graph attains the
appearance shown on Fig.~\ref{fig:DPL-ideal}. Although each node now
carries three labels the microscopic vertex weight $\lambda({\bf x})$
remains a single-valued
function on ${\cal I}$. The repeat lattice is ${\cal R} = \Zs$, and the
electric charges live on the reciprocal lattice ${\cal R}^* = 2\pi \Zs$.
For the background electric charge we find the unique solution
${\bf e}_0 = \pi e$, where $n = 2 \cos(\pi e)$ is the loop fugacity.
Since ${\cal R}_w^* = {\cal R}^*$ there are two candidates for the
screening charges, ${\bf e}_w^{(1)} = 2\pi$ and ${\bf e}_w^{(2)} = -2\pi$.
The former is associated with the most relevant vertex operator, and
the loop ansatz fixes the (unique) elastic constant appearing in $S_{\rm E}$
to be $K = \frac{\pi}{2}(1-e)$. Finally, the central charge
and the geometrical scaling dimensions are computed following the standard
procedure, and as expected they are found to coincide with those of
dense polymers. On the other hand, if a charge asymmetry evolves
${\bf e}_w^{(2)}$ is the correct screening charge, and the corresponding
solution for the elastic constant $K = \frac{\pi}{2}(1+e)$ is found to
lead to the critical exponents of dilute polymers.

It should now be clear that our ideas generalise to an arbitrary lattice
of fixed coordination number $z$, provided that the usual Coulomb gas
construction can be carried out. The description of the compact phase
procedes via a $z$-colouring description, and the colour vectors
${\bf C}_i$ must satisfy $\sum_{i=1}^z {\bf C}_i = 0$. Accordingly, the
continuum field theory contains $z-1$ massless bosons, which in the presense
of a background electric charge will couple in some non-trivial way.
Now consider allowing violations of the fully-packing constraint. This
will cause the $[z/2]$ pairs of colour vectors, each defining one flavour
of loops, to sum to zero separately, and, if $z$ is odd, the remaining colour
vector to vanish. The repeat lattice will then be a $[z/2]$-dimensional
hypercube, and the loop flavours will decouple in the familiar way.
Of the $z-1$ original bosons $[(z-1)/2]$ become massive, whilst the
remaining $[z/2]$ describe each one independent flavour of dense or
dilute polymers.

A final remark pertains to our observation that in the DPL$^2$ model
one of the two loop flavours may flow to the dilute phase. Our Coulomb
gas construction naturally suggests that it should also be possible for
{\em both} flavours simultaneously to be dilute. However, it is not obvious
that this is compatible with the vertex configurations of
Fig.~\ref{fig:vertices}, since the fractal dimension $D_{\rm f} =
2-x_2 = (3+2e)/(2+2e)$ of dilute polymers 
is always less than two.%
\footnote{Of course the dimension of the {\em union} of all loops
generated by the vertices of Fig.~\ref{fig:vertices} is always two,
regardless of the loop fugacities, but in the limit
$(n_{\rm b},n_{\rm g}) \to (0,0)$ where there is only a {\em single} loop
of either flavour there is still a discrepancy.}
On the other hand, it is clear that the Coulomb
gas construction of Sec.~\ref{sec:DPL2} would have remained exactly the
same if we had allowed some additional DPL$^2$ vertices with two or four
of the edges being unoccupied by any of the two loop flavours. Evidently,
in this generalised model loops would be allowed to have any fractal
dimension, $1 \le D_{\rm f} \le 2$.

The authors would like to thank B.~Duplantier, B.~Nienhuis and
H.~Saleur for useful
discussions and the organisers of the 1998 Les Houches Summer School
on {\em Topological aspects of low-dimensional systems} for providing
a stimulating environment during which this work was conceived. JLJ
furthermore acknowledges hospitality at {\sc Nordita}, and JK
financial support through NSF grant number DMS 97-29992.

\end{multicols}


\begin{references}

  \bibitem{Nienhuis}      B.~Nienhuis, in {\em Phase Transitions and Critical
                          Phenomena}, edited by C.~Domb and J.~L.~Lebowitz
                          (Academic, London, 1987), Vol.~11.

  \bibitem{nien_fpl}      H.~W.~J.~Bl\"{o}te and B.~Nienhuis,
                          Phys.~Rev.~Lett.~{\bf 72}, 1372 (1994).

  \bibitem{batch94}   	  M.~T.~Batchelor, J.~Suzuki and C.~M.~Yung,
			  Phys.~Rev. Lett.~{\bf 73}, 2646 (1994).

  \bibitem{jk_jpa}        J.~Kondev, J.~de Gier and B.~Nienhuis,
                          J.~Phys.~A {\bf 29}, 6489 (1996).

  \bibitem {batch96}      M.~T.~Batchelor, H.~W.~J.~Bl{\"o}te, B.~Nienhuis
                          and C.~M.~Yang, J.~Phys.~A {\bf 29}, L399 (1996).

  \bibitem{jj_npb}        J.~L.~Jacobsen and J.~Kondev, cond-mat/9804048.
                          To appear in Nucl.~Phys.~B.

  \bibitem{DupSaleur87}   B.~Duplantier and H.~Saleur,
                          Nucl.~Phys.~B {\bf 290}, 291 (1987). 

  \bibitem{chan_dill}     H.~S.~Chan and K.~A.~Dill,
                          Macromolecules {\bf 22}, 4559 (1989).

  \bibitem{orland}        H.~Orland, C.~Itzykson and C.~de Dominicis,
                          J.~Phys.~(Paris) {\bf 46}, L353 (1985).

  \bibitem{Blote89}       H.~W.~J.~Bl\"{o}te and B.~Nienhuis,
                          J.~Phys.~A {\bf 22}, 1415 (1989).

  \bibitem{jk_prl}        J.~Kondev, Phys.~Rev.~Lett.~{\bf 78}, 4320 (1997).

  \bibitem{jk_prb}        J.~Kondev and C.~L.~Henley,
                          Phys.~Rev.~B {\bf 52}, 6628 (1995).

  \bibitem {KT}           J.~M.~Kosterlitz and D.~J.~Thouless,
                          J.~Phys.~C {\bf 6}, 1181 (1973).

  \bibitem{jk_npb}        J.~Kondev and C.~L.~Henley,
                          Nucl.~Phys.~B {\bf 464}, 504 (1996).

  \bibitem{Dotsenko}      Vl.~S.~Dotsenko and V.~A.~Fateev,
                          Nucl.~Phys.~B {\bf 240}, 312 (1984);
                          {\em ibid.} {\bf 251}, 691 (1985).

  \bibitem{jk_unpub}      J.~Kondev, unpublished.

  \bibitem{fuchs}         J.~Fuchs, {\em Affine Lie Algebras and 
                          Quantum Groups}, (Cambridge University Press, 
                          Cambridge, 1992). 

\end{references}
\end{document}